# Optical versus electron diffraction imaging of Twist-angle in 2D transition metal dichalcogenide bilayer superlattices

S. Psilodimitrakopoulos[1], A. Orekhov[2,3], L. Mouchliadis[1,5], D. Jannis[2,3], G.M. Maragkakis[1,3], G. Kourmoulakis[1,5], N. Gauquelin[2,3], G. Kioseoglou[1,5], J. Verbeeck[2,3], E. Stratakis[1,4,5].

[1]Institute of Electronic Structure and Laser, Foundation for Research and Technology-Hellas, Heraklion Crete 71110, Greece.

[2] Electron Microscopy for Materials Science (EMAT) University of Antwerp 2020 Antwerp, Belgium.

[3]NANOlab Center of Excellence, University of Antwerp, Belgium.

[4]Department of Physics, University of Crete, Heraklion Crete 71003, Greece.

[5]Department of Materials Science and Technology, University of Crete, Heraklion Crete 71003, Greece.

Correspondence: S. Psilodimitrakopoulos; E. Stratakis, E-mail: sopsilo@iesl.forth.gr; stratak@iesl.forth.gr

**Abstract**

Atomically thin two-dimensional (2D) materials can be vertically stacked with van der Waals bonds, which enable interlayer coupling. In the particular case of transition metal dichalcogenide (TMD) bilayers, the relative direction between the two monolayers, coined as twist-angle, modifies the crystal symmetry and creates a "superlattice" with exciting properties. Here, we demonstrate an all-optical method for pixel-by-pixel mapping of the twist-angle with resolution of 0.23(°), via polarization-resolved second harmonic generation (P-SHG) microscopy and we compare it with four-dimensional scanning transmission electron microscopy (4D-STEM). It is found that the twist-angle imaging of $WS_2$ bilayers, using the P-SHG technique is in excellent agreement with that obtained using electron diffraction. The main advantages of the optical approach are that the characterization is performed on the same substrate that the device is created on and that it is three orders of magnitude faster than the 4D-STEM. We envisage that the optical P-SHG imaging could become the gold standard for the quality examination of TMD superlattice-based devices.

**Introduction**

Following the discovery of graphene, the appearance of 2D transition metal dichalcogenides (TMD) significantly broadened the knowledge in the field of 2D materials, as well as opening potential optoelectronic applications. Besides this, vertical stacks of two TMD monolayers (ML) forming a bilayer demonstrate exciting optoelectronic properties, not present in individual MLs [1]. In such TMD bilayers, the two constituent MLs may possess different crystal directions creating in their overlapping region a moiré superlattice. The relative direction between the two MLs is called the twist-angle.

Photoluminescence (PL) studies on $MoS_2$ TMD bilayers with different twist-angles reveal an interlayer electronic coupling, which corresponds to an indirect bandgap recombination which varies with twist-angle [2]. Recently, it was also discovered that the moiré periodic potential in twisted $MoS_2$ bilayer can modify the properties of phonons in the respective ML constituents to generate Raman modes related to moiré phonons [3]. More recent studies revealed the presence of moiré excitons in twisted TMD homo- and hetero-bilayers [4-6]. The moiré pattern in the crystal symmetry of a twisted bilayer can be controlled through the rotation of the adjacent layers. In this context, the twist-angle is regarded as a new degree of freedom, enabling tuning of the physical properties of the TMD superlattices. Consequently by tuning the twist-angle in real-space, the change in the moiré pattern and consequently to the moiré periodic potential, one can control the interlayer coupling in order to obtain the desired superlattice properties. Obviously, the precise characterization of the twist-angle in moiré superlattices is essential for a global understanding and quality control in such 2D material systems, as well as precise tuning of the respective vdW devices' performance.

Although TEM is the most commonly used technique to atomically reconstruct twisted TMD bilayers [7], it requires tedious sample transfer on TEM grids, which is incompatible with most 2D materials fabrication and characterization techniques. Apart from being technically challenging, this procedure might eventually distort the relative lattice direction and alignment. Twisted TMD bilayers have also been imaged using atomic force microscopy (AFM) [8], but this requires direct contact with the active area of the heterostructure, thus excluding the use of hBN or other forms of encapsulation. While scanning electron microscopy (SEM) techniques do not generally suffer from these limitations, conventional SEM techniques used for crystallographic imaging rely on detection of backscattered primary electrons, which is not enough to probe mono- or bilayer materials [9].

As far as the optical techniques are concerned, up to date, the estimation of twist-angle in 2D TMD bilayers has been based either on simple optical microscopy observations, or on the production of SHG signals [5,6]. However, the approaches reported to date do not exhibit high accuracy and cannot image the twist-angle over extended bilayer areas. Having a tool that spatially resolves, with high-resolution and minimally-invasively, the twist-angle in large area vdW heterostructures, would be therefore of great importance in the quality characterization of such structures. In this work we demonstrate such a technique based on the areal imaging of polarization-resolved SHG (P-SHG) signals from TMD superlattices complemented with theoretical modelling that predicts the SHG signals interference from the respective overlapping areas of twisted-bilayers.

2D TMD materials like $WS_2$ MLs belong to the $D_{3h}$ point symmetry group with broken inversion symmetry along the armchair direction. This lack of inversion symmetry in the TMD ML results in coherent SHG signals, when an intense field is incident on the 2D material [10,11]. The non-centrosymmetry which creates the SHG signals originates from the honeycomb lattice of the $WS_2$ 2D crystal, because of the alternating S and W atoms (top view in bilayer atoms configuration of Fig. 1a). The positions of the alternating S and W atoms define the broken symmetry axis of the crystal, which lies in the armchair direction. In contrast, the alternating S and S or W and W atoms define the zig-zag direction of the crystal.

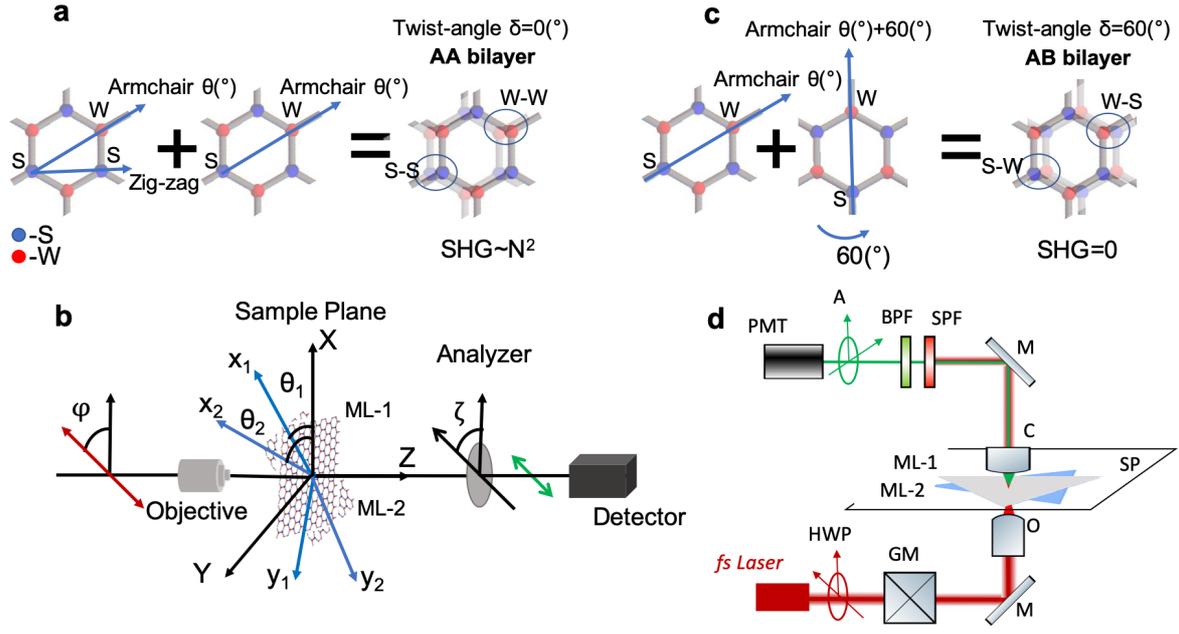

**Fig. 1 SHG signals originating from $WS_2$/ $WS_2$ bilayers a,** Schematic representation of the top view in the atomic configuration of a 2D $WS_2$ bilayer, for the AA stacking sequence. Atoms of W-W or S-S are on top of each other and the SHG signals depend quadratically on the number N of the MLs, (SHG~$N^2$). **b,** Coordinate system of the experimental configuration used for P-SHG imaging. $\theta_1$ and $\theta_2$ denote the armchair directions of ML-1 and ML-2, respectively. $\varphi$ indicates the direction of the excitation linear polarization and $\zeta$ the axis of the analyzer. In our experiments $\zeta$ is fixed to 0(°) which greatly simplifies the experimentally retrieval of the P-SHG polar diagrams. **c,** Atomic configuration for the AB stacking sequence. Alternating W-S atoms are on top of each other and the SHG signals cancel (SHG=0), because centrosymmetry is restored. **d,** Block diagram of the experimental setup used for P-SHG imaging. HWP: half-wave plate, GM: galvanometric mirrors, M: mirror, O: objective, SP: sample plane, C: condenser, SPF: short pass filter, BPF: band-pass filter, A: analyzer, PMT: photomultiplier tube.

The generated SHG from a $WS_2$ ML (crystal class $D_{3h}$) is described by its corresponding susceptibility tensor, $\chi^{(2)}$. In our approach, we rotate the direction of the linear polarization $\varphi$ of the excitation field and we detect the SHG component parallel to X-axis ($\zeta=0$(°) in Fig. 1b). Then, the recorded SHG from a $WS_2$ ML, is given by (a "four leaved rose"-like, polar diagram) [12]:

$$I^{2\omega}_{(ML)} = [A\cos(3\theta_1 - 2\varphi)]^2. \qquad (1)$$

Here $A=E_0^2\varepsilon_0\chi_{xxx}^{(2)}$, with $\varepsilon_0$ being the dielectric constant, $E_0$ the amplitude of the excitation field and $\theta_1$ [0(°)–60(°)], defines the armchair direction of the ML modulo 60(°). This means that the armchair directions with $\theta_1(°)+k60(°)$ in Eq. (1) (where k is an integer) will provide the same P-SHG polar diagram.

When two TMDs, i.e. two WS$_2$ MLs, are stacked to form a bilayer, the SHG signals from each ML interfere and the total SHG intensity from the WS$_2$/WS$_2$ superlattice is described by:

$$I^{2\omega}_{(BI)} = A^2[cos(3\theta_1 - 2\varphi) + cos(3\theta_2 - 2\varphi)]^2 \tag{2}$$

Now, one can use the concept of effective armchair direction $\theta_{eff}$ in the overlapping region of the two WS$_2$ monolayers and express the total SHG intensity produced by the 2 MLs as [7].

$$I^{2\omega}_{(BI)} = [A_{eff}cos(3\theta_{eff} - 2\varphi)]^2 \tag{3}$$

where

$$A_{eff} = 2Acos\frac{3}{2}\delta, \tag{4}$$

where $\delta=\theta_1-\theta_2$ is the twist-angle, between the MLs and

$$\theta_{eff} = \frac{\theta_1+\theta_2}{2}. \tag{5}$$

As a result, the P-SHG modulation emerging from a bilayer region consisting of two WS$_2$ MLs, at twist angle $\delta=\theta_1-\theta_2$, behaves as if it was the P-SHG modulation of a single ML with armchair direction $\theta_{eff}$. The $\theta_{eff}$ can be extracted experimentally from the P-SHG polar obtained from the bilayer region. In the above, the excitation field propagates along Z-axis and the $x_1$ armchair direction of WS$_2$ ML-1 is at angle $\theta_1$ with respect to X-axis (Fig. 1b). While the second WS$_2$ ML-2 has its armchair direction lying in $x_2$ direction, at angle $\theta_2$ with respect to X-axis (Fig. 1b).

Note in Eq. (4), that the SHG intensity from the twisted bilayer depends on the twist-angle $\delta$, being maximum for $\delta=0(°)$ and zero for $\delta=60(°)$. When the two WS$_2$ MLs are perfectly aligned, e.g. for ($\delta=0(°)$), we have the AA stacking sequence (S or W atoms in one layer lie respectively above the S or W atoms of the second layer (S-S, W-W in Fig. 1a). In this case, the total SHG signal from the bilayer is the result of constructive interference, resulting in SHG intensity four times larger than that of the ML (for N number of MLs the produced SHG signal is analogous to $N^2$). In the case of AB stacking sequence ($\delta=60(°)$) the S or W atoms in one layer lie respectively above the W or S atoms of the second layer (S-W, W-S in Fig. 1c). In this case centrosymmetry is restored and the SHG signal from the bilayer vanishes (SHG=0).

The above considerations, refer to the ideal cases of complete constructive or destructive interference in AA and AB bilayer stacking sequences, respectively. However, deviations from these ideal stacking sequences, can occur [13]. A direct consequence of such departures from the stacking sequences AA and AB is the incomplete constructive or destructive interference of the SHG from different MLs. In this case, the P-SHG signal modulation depends on the twist-angle between the MLs as shown in Fig. 2.

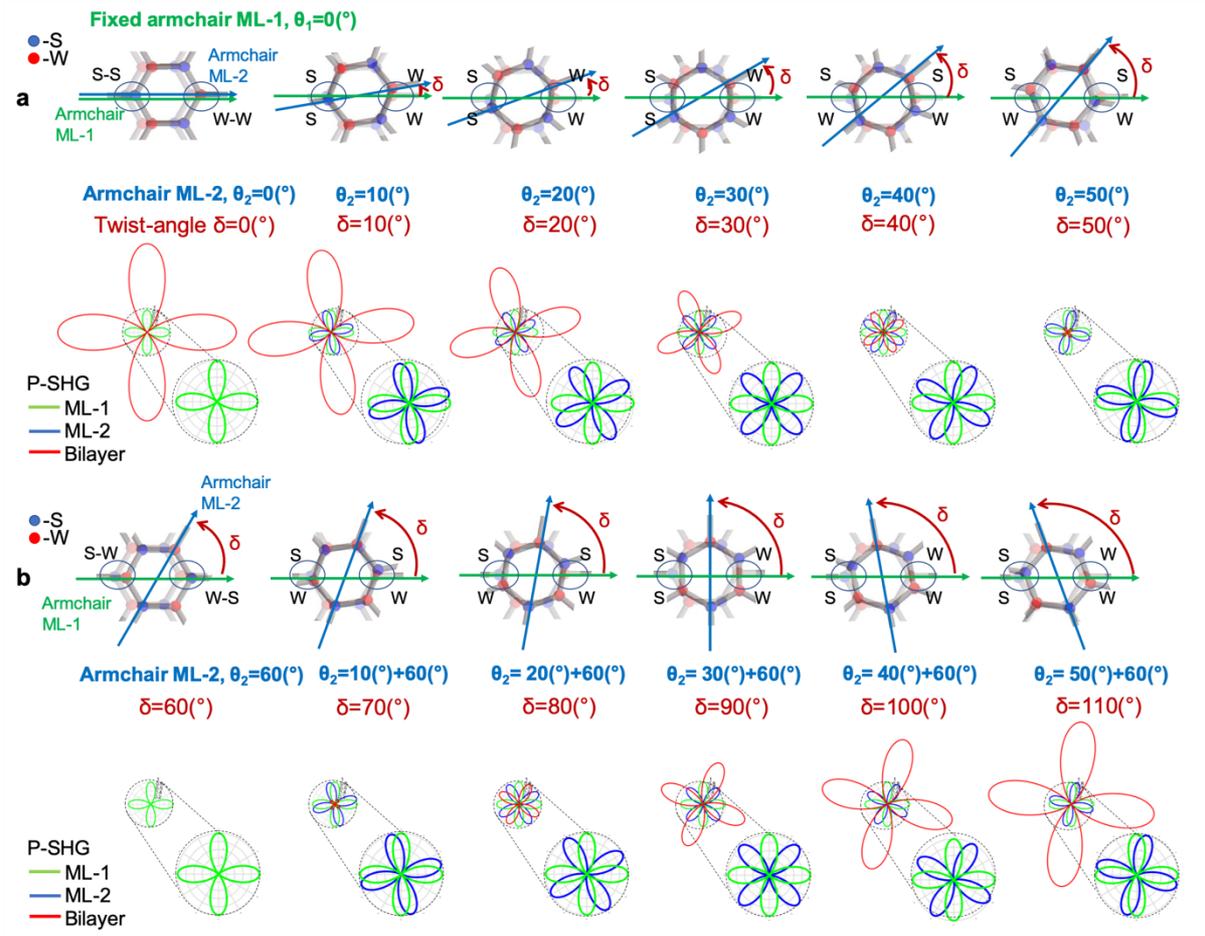

**Fig. 2 P-SHG interference signatures in twisted $WS_2/WS_2$ bilayers a,** Schematic showing the dependence of P-SHG signal on the twist-angle for fixed armchair direction ML-1, ($\theta_1=0(^0)$) and varying armchair direction $\theta_2$ of ML-2, $\theta_2 \in [0(°)–50(°)]$, with step 10(°). Top view atom configuration hexagonal lattice of the bilayer and simulated polar diagrams of the interference P-SHG originated from the $WS_2/WS_2$ bilayer (in red), as well as the simulated P-SHG signal modulation of each individual $WS_2$ ML (green for ML-1 and blue for ML-2). **b,** P-SHG signal modulation for fixed $\theta_1=0(^0)$ and varying $\theta_2 \in [60(°)–110(°)]$, with step 10(°). By comparing a with b, we notice that the same green and blue P-SHG polar diagrams can provide two different interference P-SHG red polar diagrams.

Therefore, we are able to define unequivocally the armchair direction $\theta_2$ that contributes to the SHG signal of the bilayer in the range of $0(°) \leq \theta_2 \leq 120(°)$.

The graphical representation of Eq. (1) and the corresponding visualization in a polar diagram (presented in Fig. 2) demonstrates a fourfold symmetry of the P-SHG intensity modulation that rotates for different armchair directions θ. Thus, each armchair direction corresponds to a characteristic fourfold symmetric ("four-leaved rose" like) polar diagram. Consequently, we can calculate θ by fitting Eq. (1) to experimentally retrieved P-SHG signal intensities with analyzer parallel to X -axis and different linear excitation directions $\varphi$. The experimental configuration of a fixed analyser parallel to the X -axis used here ($\zeta=0(°)$ in Fig. 1b), is much simpler to that of a rotating analyser parallel to the rotating linear excitation polarization ($\zeta=\varphi$ in Fig. 1b), used previously [6].

The green polar diagrams in Fig. 2 correspond to $\theta_1=0(°)$ (created using Eq. (1)), while the blue polar diagrams correspond to $0(°) \leq \theta_2 \leq 110(°)$ with a step of $10(°)$ (created again using Eq. (1)). The red polar diagrams correspond to the product of interference P-SHG in the bilayer region (created using Eq. (2)).

By comparing Fig. 2a, with Fig. 2b, we note the effect of the modulo of $60(^0)$ in the calculation of the armchair direction of individual MLs. That is, individual MLs with armchair directions $\theta(^0)+k60(^0)$, where k is an integer, will provide the same P-SHG polar diagrams. This implies that P-SHG measurements can calculate the armchair direction of an individual ML in the range $0(°)$–$60(°)$. Nevertheless, in the case of the bilayer, the SHG signals originate from constructive or destructive interference due to the atomic phase matching between the individual MLs (Fig. 2). Thus, the produced P-SHG modulation from the bilayer (calculated using Eq. 2) dictates a new angle range for the armchair direction $\theta_2$ of the second ML-2, i.e. $0(°)$–$120(°)$.

## Results and Discussion

In order to create the $WS_2/WS_2$ bilayer, two $WS_2$ monolayers were produced by mechanical exfoliation and were stacked with dry stamping on a $Si_3N_4$ support-grid (see Methods). This allows direct comparison between P-SHG and 4D STEM. The excitation source used for the SHG experiments is a fs oscillator, at 1030 nm and repetition rate in the order of MHz, which is adequate to excite non-linear signals like SHG (see Fig. 1b and Methods for the P-SHG microscope).

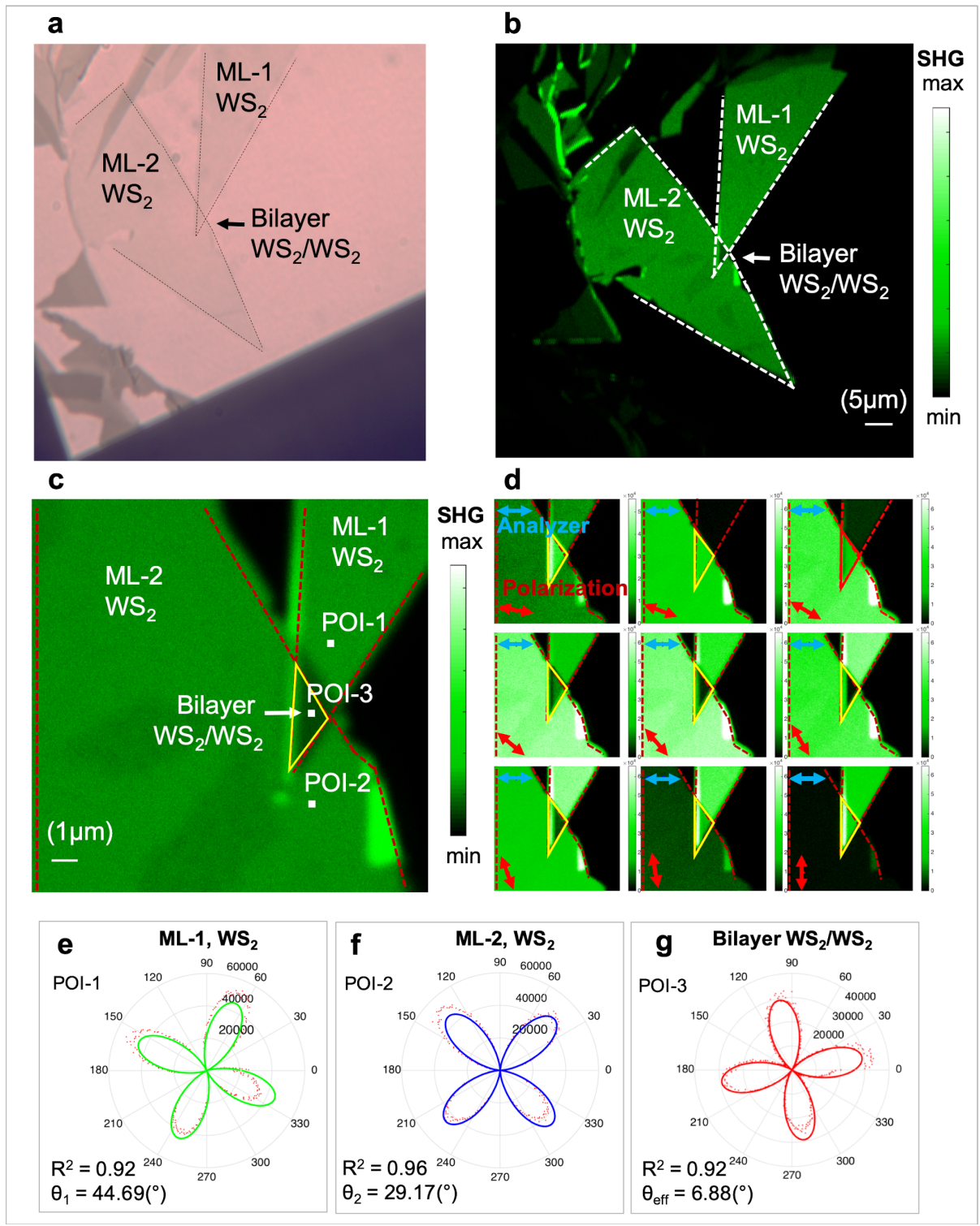

**Fig. 3 P-SHG microscopy of WS$_2$/WS$_2$ bilayer on a TEM grid a,** Wide field microscopy using the CCD camera of the microscope. We note two individual MLs of WS$_2$ (ML-1 and ML-2), overlapping in a bilayer WS$_2$/WS$_2$ region. **b,** SHG imaging of the same region seen in a., without the use of the analyser in the detection path (see Figs. 1b,d). This ensures that all of the produced SHG signals will be acquired in the same image. **c,** A zoom of the SHG image seen in b. Three pixels-of-interest (POI) are chosen. One from the $WS_2$ ML-1, one from the $WS_2$ ML-2 and one from the $WS_2/WS_2$ bilayer region. **d,** P-SHG images of the region seen in c. for fixed analyser ($\zeta=0(°)$), indicated by a blue double arrow and rotating linear excitation polarization ($\varphi \in [10(°)-90(°)]$, with step $10(°)$), indicated by a red double arrow. **e,f,** Experimentally retrieved P-SHG polar diagrams for the POIs 1,2, respectively. The polarization rotates with $\varphi \in [0(°)-360(°)]$, and step $1(°)$ and the analyser is fixed at $\zeta=0(°)$. The solid lines (green for ML-1 and blue for ML-2) are the fittings of Eq. (1) to the P-SHG data. This resulted to armchair directions $\theta_1=44.69(°)$, $R^2=0.92$ and $\theta_2=29.17(°)$, $R^2=0.96$ for ML-1 and ML-2, respectively, where $R^2$ denotes the quality of the fitting. **g,** Experimentally retrieved P-SHG data for POI 3 in the $WS_2/WS_2$ bilayer region using the same rotating polarization $\varphi \in [0(°)-360(°)]$, with step $1(°)$ and fixed analyzer at $\zeta=0(°)$ as in e., f. The solid red line is the fitting of Eq. (3) to the P-SHG data and $\theta_{eff}=6.88(°)$, $R^2=0.92$ is the result of the fitting.

In Fig. 3a, a CCD image of two individual $WS_2$ MLs (ML-1 and ML-2), forming a $WS_2/WS_2$ bilayer on the supporting TEM grid, is shown. One can identify two different types of regions, created by the above stacking, which produce different SHG signals (Fig. 3b,c). Specifically, we identify regions where only one of the two ML-1, and ML-2, $WS_2$ is present, as well as the region where the two MLs spatially overlap and create the $WS_2/WS_2$ bilayer. We note that, in the absence of an analyser in the detection path, the SHG intensity from the bilayer region is lower than the SHG from the MLs regions.

In order to calculate the twist-angle of the bilayer, we utilize high-resolution P-SHG measurements with a step of $\varphi=1(°)$. Fig. 3d presents the respective P-SHG images for fixed analyser ($\zeta=0(°)$) and varying direction of the excitation linear polarization for $\varphi \in [0(°)-360(°)]$. The corresponding polar diagrams (Fig. 3e,f) obtained from two points of interest (POIs 1,2 in Fig. 3c), one for each ML, are fitted to Eq. (1) in order to calculate the armchair direction of the MLs. Then, by using the effective armchair $\theta_{eff}$ obtained from the bilayer (POI 3 in Fig. 3c), we can calculate the twist-angle in the superlattice (using the experimentally retrieved polar of Fig. 3g). We find that $\theta_1=44.69(°)$, $R^2=0.92$ in the ML-1 region and $\theta_{eff}=6.88(°)$, $R^2=0.92$ in the bilayer region.

In previous reports, calculation of twist-angle is performed by measuring the armchair direction of the individual MLs, outside the bilayer region and then deducing their twist angle in their overlapping area [5,6]. If we follow this strategy we obtain $\delta=\theta_1-\theta_2=44.69(°)-29.17(°)=15.52(°)$. Nevertheless, as we see in Fig. 2a, a twist-angle of $15.52(°)$ should have resulted to constructive SHG interference, i.e. the SHG signal in the bilayer region should have been more intense than the SHG signals from the MLs. This is not the case in our experimental data, where the SHG signal in the bilayer region is less intense than the individual MLs (Fig. 3e-g). Consequently, following Fig. 2b we should use the modulo 60(°) in the calculation of $\theta_1$ and obtain, $\theta_1$=44.69(°)+60(°)=104.69(°). As a consequence, we unequivocally determine the twist-angle δ=$\theta_1$-$\theta_2$=104.69(°)-29.17(°)=75.52(°). We could also use the $\theta_{eff}$ acquired from the bilayer region and Eq.(5) to retrieve the twist-angle as δ=$\theta_1$-$\theta_2$=2($\theta_1$-$\theta_{eff}$)=2(44.69(°)-6.88(°))=75.62(°)

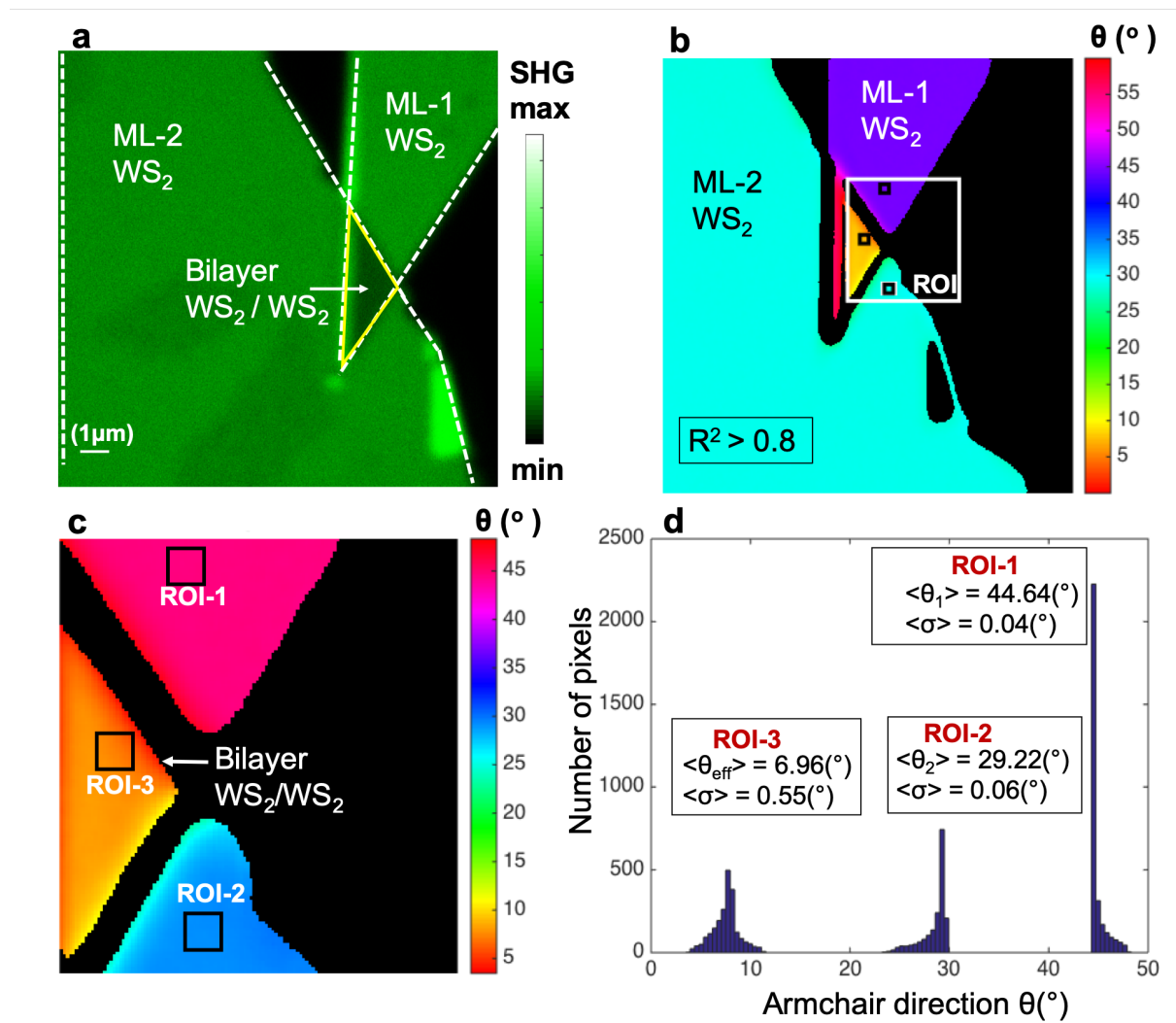

**Fig. 4 Optical imaging of twist-angle in WS$_2$/WS$_2$ bilayer. a,** SHG image without the use of analyser, showing the two ML-WS$_2$ regions and the bilayer WS$_2$/WS$_2$ region. We note that the bilayer region has less SHG signal than the MLs regions. **b,** Mapping of the armchair direction, obtained via pixel-by-pixel fitting of 360 P-SHG images, one for every $\varphi$∈[0(°)–359(°)], with step 1(°) and fixed analyzer (ζ=0(°)) to Eq. (1). Only pixels that presented R$^2$>0,8 are kept. **c,** Pixel-by-pixel mapping of the armchair direction for the ROI seen in b. Three ROIs are defined. ROIs-1,2 for the WS$_2$ MLs and ROI-3

inside the bilayer WS$_2$/WS$_2$ region. **d,** Image histogram showing the distribution of armchair directions seen in c. The mean armchair directions and their standard deviation (σ) of the armchair values found in ROIs1-3 are presented.

In Fig. 4(b) we present pixel-wise mapping of the armchair direction for the region seen in Fig. 4(a). We used 360 P-SHG images, each one acquired for 1(°) rotating excitation linear polarization for $\varphi \in [0(°)–359(°)]$ and we have fitted them pixel-by-pixel to Eq. (1). We choose to keep only the pixels that exhibit a quality of fitting, $R^2>0.8$. As we have explained in [12] the P-SHG signal fails to fit in the borders among the regions of different armchair orientations in Figs. 4b,c. As a consequence the pixels located at these borders are missing from the armchair map.

In Fig. 4(c) we present a magnification of the ROI marked in Fig. 4(b). We have also marked two ROIs-1,2, one for each ML-1,2, respectively and ROI-3 in the bilayer region. In Fig. 4d we show the image histogram of the armchair directions present in Fig. 4c. In the same Fig. 4d we present the mean and the standard deviation (σ) values of the armchair directions found in ROIs 1-3. We found $<\theta_1>=44.64(°)$ with σ=0.04(°), $<\theta_2>=44.64(°)$, σ=0.06(°) and $<\theta_{eff}>=6.96(°)$, σ=0.55(°). If we follow the strategy of deducing the twist-angle by measuring the armchair direction of the individual MLs we acquire δ=44.64(°)-29.22(°))=15.42±0.03(°). Nevertheless, as we see in Fig. 2 such twist-angle value should have resulted in constructive SHG interference, i.e. the SHG signal in the bilayer region should have been bigger than the SHG signal of the MLs (Fig. 3e-g). This is not the case for our experimental data, therefore we should follow Fig. 2 and use the modulo 60(°) in the calculation of $\theta_1$ and obtain, $\theta_1$=44.64(°)+60(°)=104.64(°). This enables us to unequivocally determine the twist-angle δ=$\theta_1$-$\theta_2$=104.64(°)-29.22(°)=75.42±0.03(°). We could also use the mean $\theta_{eff}$ acquired from ROI-3 in the bilayer region and Eq.(5) to retrieve the twist-angle as, δ=$\theta_1$-$\theta_2$=2($\theta_1$-$\theta_{eff}$)=2(44.64(°)-6.96(°))=75.36±0.23(°).

4D STEM was employed to measure the crystal directions of the individual monolayers, as well to independently estimate the twist-angle in the WS$_2$/WS$_2$ bilayer region seen in Fig. 3a. The schematic representation of the application of 4D STEM method for studying WS$_2$ monolayers is shown in Fig. 5a. The microscope settings are presented in the Methods section. In particular, the bilayer region that has been analysed by P-SHG is scanned with 256x256 probe positions with a step size of 25 nm. A virtual dark field image (VDF), calculated by summing the intensity of diffraction spots in each probe position over the selected virtual aperture, is shown in Fig 5b. The inner and outer radii of the virtual aperture are chosen to select the second order diffraction spots of the $WS_2$ structure (Fig. 5c). From the VDF intensity distribution, we can identify four distinct regions corresponding to the silicon nitride support-gird (dark contrast) (mean diffraction pattern is shown in Fig. 5c), two monolayers (Fig. 5b and Fig. 5e) and an overlapped region with a bilayer (Fig. 5f). At the edges of the monolayers there are regions with increased intensity due to their folding (white arrows in Fig. 5b).

To calculate a direction map of the region,a custom-made peak finding routine was used. Once peak positions are found, two reciprocal vectors can be fitted to describe all diffraction spots defining the relative crystal direction in every probe position. This algorithm works only for diffraction patterns from monolayers. For the bilayer region, four vectors are fitted to each acquired pattern when more than six peaks are detected inside the virtual aperture area. The relative direction map of region with a bilayer area (white square in Fig. 5b) is shown in Fig. 6a. The tilt angle of every diffraction pattern for two sheets is presented in the histogram Fig. 6d. The twist-angle between the two monolayers was calculated with subpixel accuracy by fitting a Gaussian function for each peak on the histogram and was found to be 15.5±0.3°. These results are in good agreement with the experiments on the same area using the all-optical P-SHG imaging microscopy technique, presented above.

However, for the direction determination only the positions of the diffraction spots were used without taking into account their intensity. By looking at the diffraction spot symmetry, one might think that $WS_2$ crystal have six-fold symmetry along the [001] direction, however, due to dynamic electron scattering in combination with the noncentrosymmetry of the crystal, Friedel's law is violated resulting in a symmetry reduction to three-fold symmetry and a nonequivalent angular range from 0° to 120° instead of 0° to 60°. Fig. 6c and Fig. 6d show diffraction patterns of the first and second monolayer together with their azimuthal intensity profiles of the first order spots (Fig. 6e) clearly showing 3 fold symmetry. This allows to identify whether the twist angle is 15.5° or 75.5° (15.5°+60°) by comparing to simulated diffraction patterns shown in Fig. 6f,g. The simulation is performed using the Multem software [14,15] where the experimental conditions such as convergence angle and electron beam energy are chosen to closely resemble the experimental setup (see Methods section). A noticeable difference between the simulated and experimental patterns (Fig. 6h) is caused by the presence of the $Si_3N_4$-support film (experimental), which gives rise to an isotropic background signal. As can be seen in the simulated patterns the intensity of the first order spots changes azimuthally from two low, two high for the twist angle 15.5°, while it has alternating low-high character for the 75.5° angle. The annular intensity profile of the first order spots of the experimental pattern (Fig. 6k), shows a clear signature of a 75.5±0.3° twist angle and rules out the 15.5±0.3° hypothesis. This result is in a excellent agreement with the P-SHG data presented above.

In terms of precision, the STEM data demonstrates 0.3° rotational precision compared to 0.55° for the P-SHG optical result with a significantly higher spatial resolution but with the downside of a technique which is far less attractive to employ in an inline production environment as compared to the all optical setup proposed in this paper.

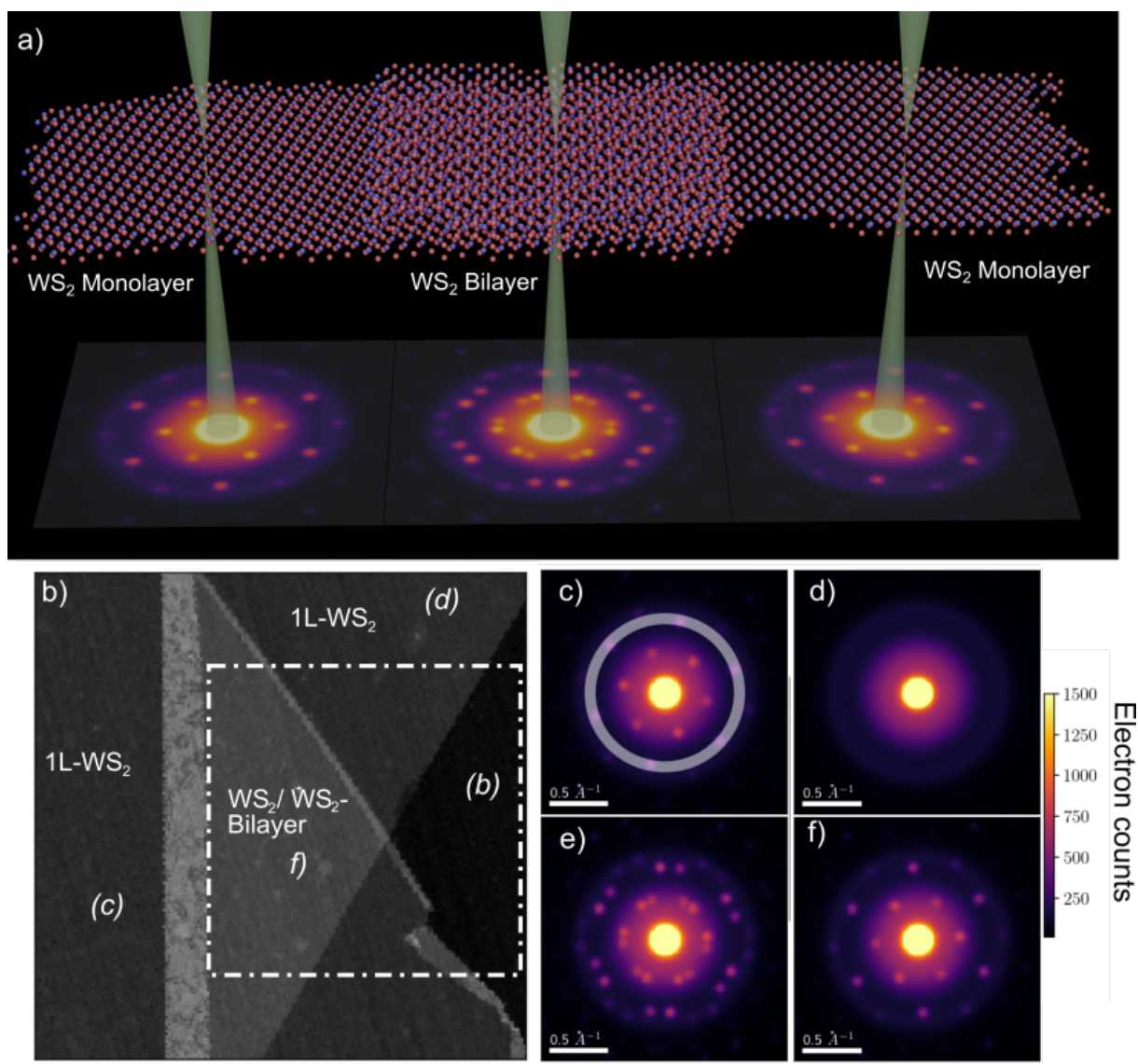

**Fig 5 4D STEM measurements. a** Schematic representation of the application of the 4D STEM method for studying WS$_2$ monolayers. **b** Virtual dark field image where the intensity in each probe position is summed over selected virtual aperture indicated in **c**. **c-f** The mean diffraction from the four regions - the first WS$_2$ monolayer **c**, silicon nitride supporting film **d**, the second monolayer **e** and the bilayer region **f**.

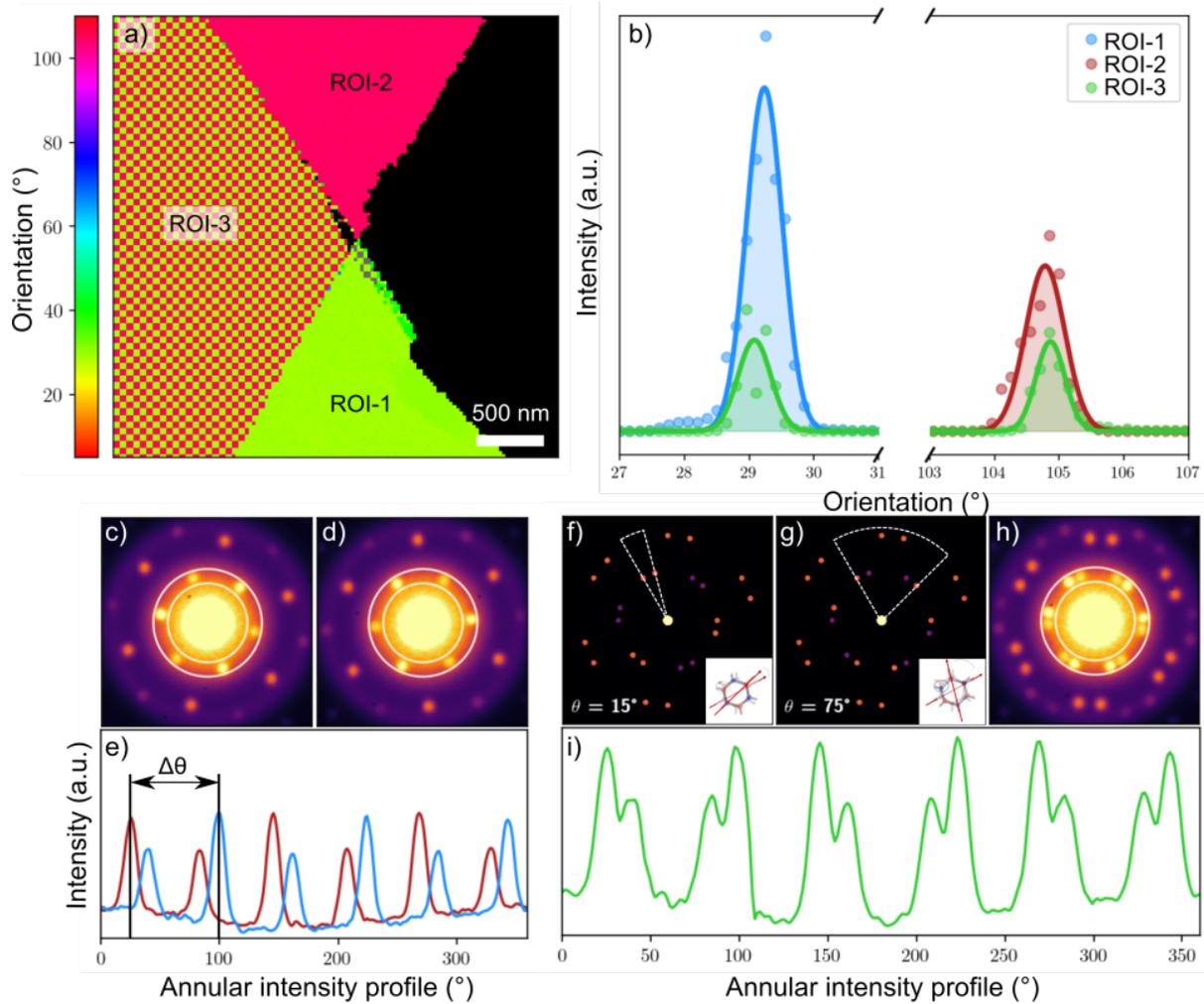

**Fig 6. Twist-angle calculation. a** The relative color-coded direction map of a bilayer region selected in Fig. 6b. **b** Histogram of the relative direction of the monolayers and the bilayer indicating that the two layers are stacked upon each other since its direction in the bilayer correspond to both monolayers. **c-d** The diffraction pattern averaged over 400 probe position for the first and second monolayer. **e** The annular intensity profile of the first order spots where a modulation of intensity is observed. **f-g** The simulated diffraction patterns of the $WS_2$ bilayer when the twist-angle is 15.5° and 75.5° (60°+15.5°). **h** Experimental diffraction of the bilayer region, averaged over 400 probe positions. **i** The annular intensity profile of the first order spots of experimental diffraction where the intensity modulation of the two layers is reversed indicating the twist angle of 75.5° instead of 15.5°

## Conclusions

Considering that the electronic and optical properties of 2D TMD bilayers can be tuned by changing their twist-angle, a robust and minimally-invasive tool that can provide spatially-resolved determination of the twist-angle, would be of great importance in research, production and large-scale characterization of 2D TMD bilayers. Here, we have used all-optical, laser raster-scanning, P-SHG imaging microscopy to precisely map the twist-angle in large areas of overlapping $WS_2/WS_2$ stacked monolayers and we benchmarked the results against 4D STEM electron microscopy. It is found that the twist-angle of $WS_2/WS_2$ bilayer obtained using P-SHG mapping is in excellent agreement, either in value and in precision, with that obtained using 4D STEM. It is additionally revealed that, given that the produced SHG signal from a bilayer is the vectorial addition of the SHG signals of the individual monolayers, the intensity modulation of the P-SHG signal can be used to deduce unequivocally the armchair direction. This is also in excellent agreement with 4D STEM microscopy analysis.

While STEM provides significantly higher spatial resolution, the P-SHG compensates for this with a wide range of advantages including: no need for vacuum, wide field of view, rapid data acquisition, significantly lower cost and instrument size, and most importantly its capability to work on TMD bilayers deposited on substrates without the need to transfer the films to a TEM grid. Our setup provides an accurate and robust all-optical twist-angle mapping of 2D TMD bilayers. Importantly, the technique is non-destructive, paving the way or directly correlating local twist-angle values with electronic properties, which is crucial for the development and scaling up of vdW bilayer devices with precisely controlled functionality.

**Methods**

**$WS_2/WS_2$ bilayer fabrication on a TEM grid**

Polydimethylsiloxane films (PDMS) were fabricated from 10:1 mixing ratio (SYLGARD 182 Silicone Elastomer Kit) with heat cure at 80°C for two hours. High quality $WS_2$ bulk crystals (HQ Graphene) were mechanically exfoliated directly on the aforementioned PDMS films. The films were placed on typical microscope glass slides using standard protocol [16,17]. Monolayers of these crystals were realized under an optical microscope. In order to produce the bilayer, at first, a glass slide with a $WS_2$ monolayer was mounted on a XYZ micromechanical stage under a coaxially illuminated microscope and transferred on a silicon nitride ($Si_3N_4$) support-grid using viscoelastic stamping [18]. Finally, another $WS_2$ monolayer was stamped on the previous one in a partial overlapping manner allowing for comparative P-SHG and 4D STEM imaging.

**Custom-built P-SHG microscope**

SHG imaging was performed in the forward-detection geometry using a custom-built laser raster-scanning microscope [7,11,12]. As shown schematically in Fig. 1d, a diode-pumped Yb:KGW fs oscillator (1027 nm, 90 fs, 76 MHz, Pharos-SP, Light Conversion, Lithuania), was inserted into a modified, inverted microscope (Zeiss Axio Observer Z1, Germany), after passing through a pair of silver coated galvanometric mirrors (6215H, Cambridge Technology, UK). A motorized rotation stage (M-060.DG, Physik Instrumente, Karlsruhe, Germany), holding a zero order λ/2 wave plate (QWPO-1030-10-2, CVI-Laser, USA) was used to rotate the direction of the excitation linear polarization. Then, the beam was reflected on a silver mirror at 45° (PFR10-P01, ThorLabs, Germany), placed at the turret box of the microscope, just before the objective (Plan Apochromat 40x 1.3NA, Zeiss, Germany). The SHG signals were collected from a high-numerical aperture (1.4NA) condenser lens (Zeiss) and guided into a photomultiplier tube (PMT) detector (H9305-04, Hamamatsu, Japan) using another silver mirror (CM1-P01, ThorLabs, Germany) at 45(°). The polarization extinction ratio was 28:1. In front of the PMT, a home-built mount was holding a bandpass filter (FF01-514/3-25, Semrock, USA) and a short pass filter (FF01-680/SP-25, Semrock, USA), appropriate for SHG imaging. After the filters, a film polarizer (LPVIS100- MP, ThorLabs) was inserted just in front of the PMT to measure the anisotropy of the SHG signals due to the rotation of the excitation linear polarization. Coordination of PMT recordings with the galvo-mirrors movements and with all the motors, as well as the image formation, was performed using LabView (National Instruments, USA).

**4D STEM measurements**

In order to determine local information on the single layer direction and the bilayer twist angle of 2D materials, a scanning transmission electron microscope ThermoFisher Scientific (FEI) Titan X-Ant-EM was used. The electron microscope was operated in microprobe STEM mode at 300 kV at a convergence semi-angle α of 1 mrad resulting in a probe size of 1.2 nm in diameter. Local information was obtained by scanning the electron probe over the sample and acquiring an electron diffraction pattern at every probe position [19-24] on a Medipix3 hybrid pixel direct electron detector (Quantum Detectors Merlin) [25-27] with a camera length of 115 mm and exposure time of 5 ms. This detector offers a high frame rate and high efficiency enabling the detection of individual electrons without dark or read-out noise and offering 24 bit dynamic range. The diffraction data was processed by custom-made scripts based on the open-source python library *Pixstem*[28].

**Acknowledgements**

This research has been co-financed by the European Union and Greek national funds through the Operational Program Competitiveness, Entrepreneurship and Innovation, under the call European R & T Cooperation-Grant Act of Hellenic Institutions that have successfully participated in Joint Calls for Proposals of European Networks ERA NETS (National project code: GRAPH-EYE T8EPA2-00009 and European code: 26632, FLAGERA).  LM, GKo and GKi acknowledge funding by the Hellenic Foundation for Research and Innovation (H.F.R.I.) under the "First Call for H.F.R.I. Research Projects to support Faculty members and Researchers and the procurement of high-cost research equipment grant" (Project No: HFRI-FM17-3034). . GKi, S.P. and G.M.M. acknowledge funding from a research co-financed by Greece and the European Union (European Social Fund - ESF) through the Operational Programme "Human Resources Development, Education and Lifelong Learning 2014-2020" in the context of the project "Crystal quality control of two-dimensional materials and their heterostructures via imaging of their non-linear optical properties" (MIS 5050340)". J.V acknowledges funding from FWO G093417N ('Compressed sensing enabling low dose imaging in transmission electron microscopy') from the Flanders Research Fund, EU. J.V. and N.G. acknowledge funding from the European Union under the Horizon 2020 programme within a contract for Integrating Activities for Advanced Communities No 823717 – ESTEEM3. J.V. N.G. and A.O. acknowledge funding through a GOA project "Solarpaint" of the University of Antwerp.

**Author contributions**

S.P. and G.M. performed the PSHG experiments and data analysis. L.M. performed the theoretical calculations and data analysis. G.Ko. prepared and characterized the samples. A.O, D.J. and N.G. performed the TEM experiments and data analysis. All authors contributed to the discussion and preparation of the manuscript.

**Data availability**

The data that support this study are available from the corresponding authors upon request.

**Conflict of interest**

The authors declare that they have no conflict of interest.

**Supplementary information** is available for this paper

# Supplementary Material

**Twist-angle in 2D TMD heterostructures using P-SHG**

The corresponding P-SHG expressions for a MoS$_2$/WS$_2$ heterostructure are given by:

$$I^{2\omega}_{\zeta=0(HE)} = [A_1 cos(3\theta_1 - 2\varphi) + A_2 cos(3\theta_2 - 2\varphi)]^2 \text{(fixed analyzer, } \zeta=0)$$

$$I^{2\omega}_{\zeta=\varphi(HE)} = [A_1 cos3(\theta_1 - \varphi) + A_2 cos3(\theta_2 - \varphi)]^2 \text{(rotating analyzer, } \zeta=\varphi)$$

where $A_1 = E_0^2 \varepsilon_0 \chi_1^{(2)}$ and $A_2 = E_0^2 \varepsilon_0 \chi_2^{(2)}$, with $\varepsilon_0$ being the dielectric constant, $E_0$ the amplitude of the excitation field and $\chi_1^{(2)}$ and $\chi_2^{(2)}$ are the SHG susceptibility tensors for monolayer TMD-1 and monolayer TMD-2, respectively.

If we use again the concept of *effective armchair orientation* in the overlapping region of the heterobilayer we express its SHG intensity as

$$I^{2\omega}_{\zeta=0(HE)} = [A_{eff} cos(3\theta_{eff} - 2\varphi)]^2 \text{ (fixed analyzer, } \zeta=0)$$

$$I^{2\omega}_{\zeta=\varphi(HE)} = [A_{eff} cos3(\theta_{eff} - \varphi)]^2 \text{ (rotating analyzer, } \zeta=\varphi)$$

where

$$A_{eff(HE)}^2 = A_1^2 + A_2^2 + 2A_1 A_2 \cos 3\delta$$

and

$$\theta_{eff(HE)} = \frac{1}{3} \tan^{-1}\left[\frac{A_1 \sin3\theta_1 + A_2 \sin3\theta_2}{A_1 \cos3\theta_1 + A_2 \cos3\theta_2}\right]$$

As a result, the P-SHG emerging from a MoS$_2$/WS$_2$ hetero-bilayer region consisting of two different 2D TMD materials at twist angle $\delta = \theta_1 - \theta_2$, behaves as if it was the result of a single layer with armchair angle $\theta_{eff}$. The latter can be used to obtain the unknown armchair orientation of one layer, when the armchair of the other and that of the overlapping region are known, along with the corresponding intensities:

$$\theta_{2(HE)} = \frac{1}{3} \tan^{-1}\left[\frac{A_{eff} \sin3\theta_{eff} - A_1 \sin3\theta_1}{A_{eff} \cos3\theta_{eff} - A_1 \cos3\theta_1}\right]$$